\documentstyle[11pt,epsfig]{article}
\newcommand{\zp}[3]{Z.\ Phys.\ {\bf #1} (19#2) #3}
\newcommand{\zpc}[3]{Z.\ Phys.\ {\bf C#1} (19#2) #3}
\newcommand{\pl}[3]{Phys.\ Lett.\ {\bf B#1} (19#2) #3}
\newcommand{\np}[3]{Nucl.\ Phys.\ {\bf B#1} (19#2) #3}
\newcommand{\pr}[3]{Phys.\ Rev.\ {\bf #1} (19#2) #3}
\newcommand{\prd}[3]{Phys.\ Rev.\ {\bf D#1} (19#2) #3}

\newcommand{\ibid}[3]{{\it ibid.}\ {\bf #1} (19#2) #3}

\newcommand{\ga}{\gamma}
\newcommand{\aem}{\alpha_{\rm em}}
\newcommand{\ep}{\epsilon}

\newcommand\sss{\scriptscriptstyle}

\newcommand\as{\alpha_{\sss S}}         
\newcommand\pt{p_{\sss T}}

\newcommand\mt{m_{\sss T}}
\newcommand\mtt{m_{\sss T}^2}
\newcommand\mur{\mu_{\sss R}}
\newcommand\murt{\mu_{\sss R}^2}
\newcommand\muf{\mu_{\sss F}}
\newcommand\muft{\mu_{\sss F}^2}

\newcommand{\gaga}{\mbox{$\gamma\gamma$}}
\newcommand{\QQB}{\mbox{$Q\overline{Q}$}}

\def\simgt{\rlap{\lower 3.5 pt \hbox{$\mathchar \sim$}} \raise 1pt \hbox {$>$}}
\def\simlt{\rlap{\lower 3.5 pt \hbox{$\mathchar \sim$}} \raise 1pt \hbox {$<$}}

\newcommand{\beq}{\begin{equation}}
\newcommand{\eeq}{\end{equation}}
\newcommand{\bea}{\begin{eqnarray}}
\newcommand{\eea}{\end{eqnarray}}

\def\abs#1{\left| #1\right|}
\def\bentarrow{\:\raisebox{1.3ex}{\rlap{$\vert$}}\!\rightarrow}
  \newcommand{\ccaption}[2]{
    \begin{center}
    \parbox{0.85\textwidth}{
      \caption[#1]{\small{\it{#2}}}
      }
    \end{center}
    }

\catcode`\@=11
\ifx\@Hxfloat\@Hundef\else\expandafter\endinput\fi
\let\@Hxfloat\@xfloat
\def\@xfloat#1[{\@ifnextchar{H}{\@HHfloat{#1}[}{\@Hxfloat{#1}[}}
\def\@HHfloat#1[H]{%
\expandafter\let\csname end#1\endcsname\end@Hfloat
\vskip\intextsep\vbox\bgroup\def\@captype{#1}\parindent\z@
\ignorespaces}
\def\end@Hfloat{\egroup\vskip \intextsep}
\def\section{\@startsection{section}{1}{\z@}{3.5ex plus 1ex minus .2ex}
{2.3ex plus .2ex}{\large\bf}}
\def\thesection{\arabic{section}.}

\def\appendix{\setcounter{section}{0}
 \def\thesection{APPENDIX \Alph{section}:}
 \def\theequation{\Alph{section}.\arabic{equation}}}

\def\@citex[#1]#2{\if@filesw\immediate\write\@auxout{\string\citation{#2}}\fi
  \def\@citea{}\@cite{\@for\@citeb:=#2\do
    {\@citea\def\@citea{,\penalty\@m}\@ifundefined
       {b@\@citeb}{{\bf ?}\@warning
       {Citation `\@citeb' on page \thepage \space undefined}}%
\hbox{\csname b@\@citeb\endcsname}}}{#1}}

\def\citer{\@ifnextchar [{\@tempswatrue\@citexr}{\@tempswafalse\@citexr[]}}
\def\@citexr[#1]#2{\if@filesw\immediate\write\@auxout{\string\citation{#2}}\fi
  \def\@citea{}\@cite{\@for\@citeb:=#2\do
    {\@citea\def\@citea{--\penalty\@m}\@ifundefined
       {b@\@citeb}{{\bf ?}\@warning
       {Citation `\@citeb' on page \thepage \space undefined}}%
\hbox{\csname b@\@citeb\endcsname}}}{#1}}
\relax
\begin{document}
\thispagestyle{empty}
\rightline{
        \begin{minipage}{4cm}
        CERN-TH/99-246\\
        NIKHEF/99-020\hfill \\
        hep-ph/9908483\hfill \\
        \end{minipage}        
}
\vskip 1.5cm
\begin{center}{{\bf\Large D$^*$ PRODUCTION IN TWO-PHOTON COLLISIONS}}
\vglue 1.5cm
\begin{sc}
Stefano Frixione, Michael Kr\"{a}mer\\
\vglue 0.3cm
\end{sc}
{\it CERN, TH Division\\
CH-1211 Geneva 23, Switzerland}
\vglue 0.3cm
and
\vglue 0.3cm
\begin{sc}
Eric Laenen\
\vglue 0.3cm
\end{sc}
{\it NIKHEF Theory Group\\
Kruislaan 409, 1098SJ, Amsterdam, The Netherlands}
\end{center}
\vglue 1.7cm

\begin{abstract}
\noindent
We calculate total and differential production rates of $D^*$ mesons
in two-photon collisions at LEP2.  We include full next-to-leading
order QCD corrections, and perform an extensive study of the
sensitivity of our predictions to variations of the renormalization
scale, charm mass, photonic parton distribution set and fragmentation
function. The results are compared with recent data from LEP2.
\end{abstract}

\vfill
\leftline{
        \begin{minipage}{4cm}
        CERN-TH/99-246\\
        August 1999 \\
        \end{minipage}        
}
\vfill

\newpage

\setcounter{page}{1}

\noindent

\section{INTRODUCTION}
Charm quark production in two-photon collisions at high-energy
$e^+e^-$ colliders provides new possibilities to study the dynamics of
heavy quark production; it complements the extensive analyses that
have been carried out at fixed-target experiments and at other
high-energy colliders~\cite{FMNR-97}. In two-photon
collisions\footnote{In this paper, we will only consider the case in
which the incoming photons are on-shell.}  each of the photons can
behave as either a point-like or a hadronic particle. Consequently,
one distinguishes in such collisions direct- (both photons are
point-like), single resolved- (one photon is point-like, the other
hadron-like), and double resolved (both photons are hadron-like)
production channels. The resolved channels require the use of parton
densities in the photon, whereas the production via the direct channel
is free of such phenomenological inputs. In general, the different
channels mix when including higher orders in perturbation theory, and
thus the distinction between the direct and resolved contributions
becomes non-physical and scheme-dependent.

The mass of the heavy quark, $m_{\sss Q}\gg\Lambda_{\rm QCD}$, acts as a
collinear cutoff and sets the hard scale for the perturbative calculation 
at small heavy-quark transverse momentum $\pt$. It is thus possible to
define an all-order infrared-safe cross section for open heavy flavour
production, even at $\pt=0$. The heavy quark mass also ensures that
the separation into direct and resolved production channels is
unambiguous at next-to-leading order (NLO). In the case of
charm production, however, the heavy quark mass is not very large
with respect to $\Lambda_{\rm QCD}$, and one therefore expects large
radiative corrections in perturbative QCD.  This is indeed borne out
by NLO analyses of charm production in hadron--hadron and photon--hadron
collisions, where the leading-order result sometimes only accounts for
less than 50\% of the full next-to-leading order rate
\cite{FMNR-97}.  As will be discussed at length in what follows, the
situation in photon--photon collisions is quite different, and the
predictions are under better theoretical control, in spite of the
presence of resolved channels.

The OPAL collaboration has presented \cite{OPAL-99} new data for $D^*$
production in two-photon collisions, at (mostly) $\sqrt{s_{e^+e^-}} =
189$ GeV.  Besides the total cross section
$\sigma_{\gamma\gamma}^{D^*}$, OPAL has measured the differential rate
with respect to the $D^*$ transverse momentum,
$d\sigma_{\gamma\gamma}^{D^*} / d\,\pt^{D^*}$, and pseudorapidity
$d\sigma_{\gamma\gamma}^{D^*} / d\,\eta^{D^*}$.  Recently, also L3 has
presented results for $D^*$ production in two-photon collisions
\cite{L3-99}. 
L3, however, does not apply an antitag condition for the scattered
electrons. It is thus not obvious how the corresponding data can be
compared with a theoretical calculation where the cross section is
defined in the Weizs\"acker--Williams approximation in the standard
way.

The new LEP2 data motivated us to perform an extensive study of $D^*$
production in two-photon collisions, including all NLO QCD
corrections. We use fully differential NLO Monte Carlo programs for
all the production mechanisms, constructed in refs.~\citer{MNR-92,KL-96}.
Earlier studies either focused on an individual component of the
production~\cite{KL-96}, or had mostly HERA data in
mind~\cite{MNR-92,FMNR-94}, or did not include fragmentation
functions~\cite{DKZZ-93}. Moreover, a relatively large number of
NLO-evolved photonic parton distributions~\citer{GRV-92,GRS-99}, and
new studies on charm fragmentation functions~\citer{CG-97,NO-99} 
are available, 
some of them incorporating the most recent theoretical and experimental
findings. We shall examine in detail the sensitivity of the NLO
calculations of $D^*$ meson cross sections with respect to variations
of the renormalization scale, charm mass, photonic parton distribution
sets, fragmentation function, and others.

The paper is organized as follows. In section 2 we review very briefly
the NLO Monte Carlo programs that we used in the rest of the paper. The 
results of our study, and the comparison of our calculations with the OPAL
data, are presented in section 3. We summarize our work in section 4. The
computation of photon--photon cross sections and the issue of scheme 
dependence of the direct and resolved production mechanisms is 
discussed in the appendix.

\section{NLO MONTE CARLO PROGRAMS}

We consider the process 
\beq 
e^+ + e^- \longrightarrow e^+ + e^- + Q+\overline{Q}, 
\eeq 
where the final state positron and electron are scattered almost
collinearly to the beam line. The cross section is dominated by the
scattering of two on-shell photons, which produce the heavy quarks and 
can be written as a convolution of two Weizs\"acker--Williams spectra
\cite{WW-34} with the cross section for the process 
\mbox{$\ga\ga\to Q\overline{Q}$}. As mentioned in the introduction, besides 
the direct contribution, also resolved production channels have to be
taken into account. The cross section for the resolved
contributions factorizes into a partonic hard-scattering cross section,
convoluted with photonic light-quark and gluon densities. Notice that 
these densities grow as $\aem/\as$ at large scales, owing to the 
inhomogeneous term in the relevant Altarelli--Parisi equations;
therefore, the cross sections for the resolved processes are
formally of the same perturbative order as the cross section for 
the direct process. More details on the computation of photon--photon
rates in perturbative QCD will be given in the appendix. 

To calculate the direct and resolved contributions including the
complete NLO corrections, we use Monte Carlo
programs that are fully exclusive in all final-state
particles. We want to emphasize that our programs are
different from the usual Monte-Carlo parton-shower codes, since
they result from fixed-order QCD computations, where no collinear
approximation (and no shower evolution) has been performed. For the
direct channel, the NLO program has been constructed in~\cite{KL-96},
while for the single-resolved and double-resolved channels we use the
NLO programs constructed in refs.~\cite{MNR-92,FMNR-94}.  To check our
results we employed in addition the single-particle inclusive programs
of refs.~\cite{DKZZ-93,NDE-88,NE-89}. In our codes all final-state
kinematical quantities are available on an event-by-event basis, and it
is thus possible to calculate any infrared-safe quantity to NLO, 
if the quantity requires two or three partons for its definition, or
to LO if it requires three partons, with any final-state 
kinematical cuts matching those implemented by experiments. This
allows us to compare NLO predictions with measurements in the experimentally
visible region. As will be discussed below, cross section
extrapolation beyond the accepted region introduces large
theoretical uncertainties, at least for the case at hand, and we find
the visible cross section to be as important as the total cross
section for the comparison of data with theory. The wide
availability of flexible higher order Monte Carlo programs in which
acceptance cuts can be easily built in makes the calculation of
many types of visible cross section practical, thereby providing
the possibility of a more detailed comparison between theoretical 
and experimental results at high energy colliders.

\section{RESULTS}

We will first define a default set of input parameters and then study
parameter variations to examine the sensitivity of our theoretical
predictions. The collision energy is fixed at $\sqrt{s_{e^+e^-}} =
189$~GeV. We set the charm quark mass, defined in the on-shell
renormalization scheme, to $m_c$ = 1.5~GeV. The two-loop expression
for the QCD coupling is used, with $\Lambda^{(5)}_{\overline{\rm MS}}
= 220$~MeV, as favoured by recent global QCD fits. The default choice 
for the renormalization scale is $\murt=\mtt\equiv m_c^2 + \pt(c)^2$, 
and for the factorization scale we use $\muft=4\mtt$. Our default choice for
the photonic parton densities entering the resolved channels is the
recent NLO-evolved set GRS-HO~\cite{GRS-99}. We checked that, by
adopting the GRS-HO set defined in the $\overline{\rm MS}$ or 
DIS$_\gamma$~\cite{GRV-DISgamma}
scheme, and changing the short-distance cross sections accordingly,
only numerically negligible differences arise. Finally, for the
Weizs\"acker--Williams function we adopt the form proposed in
ref.~\cite{FMNR-93}, designed to describe an antitag condition on the
scattered electron.  Here, the antitag angle is $\theta_{\rm max} =
0.033$, as determined by the OPAL experimental setup. The Peterson et
al.\ parametrization~\cite{PSSZ-83} is used as the charm-to-$D^*$
fragmentation function, normalized according to $\int_0^1 dz
D(z,\epsilon)=1$. Our default value for the non-perturbative parameter
entering the fragmentation function will be $\ep=0.035$ (see for
example ref.~\cite{NO-99}). In order to obtain the $D^*$ kinematical
variables from those of the fragmenting charm quark, we rescale the
three space components of the parton momentum with the momentum
fraction $z$, and compute the energy according to the charm
quark mass shell condition (the choice of mass to fix the $D^*$
energy component is arbitrary, because the factorization theorem only
holds for large $\pt$, where mass effects are negligible.  We
have checked that our numerical results in the experimentally visible
region $\pt(D^*)\;\simgt\;2$~GeV are not sensitive to the specific
choice).

We have to comment on the fact that the $\Lambda_{\rm QCD}$ value we
use is not the one associated with the GRS-HO photon set. The parton
densities are correlated with $\Lambda_{\rm QCD}$~\footnote{The
correlation is due to the Altarelli--Parisi evolution equation, and 
to the fact that the largest sensitivity is to $\as g(x)$, where
$g(x)$ is the gluon density, and not to $g(x)$.}, so that, in
principle, $\Lambda_{\rm QCD}$ cannot be chosen independently of the
PDF set. However, in practice, it is known that the correlation
between $\Lambda_{\rm QCD}$ and the densities of the photon is very
mild \cite{Vogt-99}, and our choice can be considered safe, also given
the fact that other theoretical uncertainties are dominant in charm
physics. Since our aim is to compare theory with experimental data and
because the normalization of the resolved contributions is very
sensitive to the choice of $\Lambda_{\rm QCD}$, we deem it appropriate
to use an up-to-date value for $\Lambda_{\rm QCD}$. As a final comment
on our default choices, we remark that we prefer to use $\muf=2\mt$
rather than $\muf=\mt$ as the factorization scale: at very low
transverse momenta, $\muf=\mt$ would probe the parton densities at a
scale where the validity of the evolution equations is not on firm
grounds. As is customary in charm physics, we will not consider the 
variation of our results with respect to $\muf$, for the same reason.
The scale uncertainty that we will quote therefore only partially
accounts for the full theoretical error.

\subsection{TOTAL CROSS SECTIONS}

The preliminary value obtained by OPAL for the total cross section for
$D^*$ production in \gaga\ collisions at $\sqrt{s_{e^+e^-}}=189$ GeV, in
the range $2<\pt(D^*)<12~\mbox{GeV}$ and with the
cut $|\eta(D^*)|<1.5$ is
\beq
\sigma(e^+e^-\to e^+e^- D^*X)_
{[2 < \pt(D^*) < 12~{\rm GeV}; |\eta(D^*)|< 1.5]} = 
29.9 \pm 4.2~\mbox{pb}\quad \mbox{(OPAL)}.
\eeq
Taking into account the probability for a charm quark to fragment into
a $D^*$ meson (we use here $f(c\to D^{*+})=0.233\pm 0.010$, the same value
as adopted in ref.~\cite{OPAL-99}), and the
fact that OPAL counts both the heavy quark and the heavy antiquark in
their cross-section determination, we find a theoretical value of
\beq
\sigma(e^+e^-\to e^+e^- D^*X)_
{[2< \pt(D^*) < 12~{\rm GeV}; |\eta(D^*)|< 1.5]} = 17.3 \begin{array}{l}
 +5.1\\[-1.5mm] -2.9 \end{array}~\mbox{pb}\quad \mbox{(NLO QCD)}.
\eeq
The theoretical error reflects the dependence on the charm quark mass,
chosen in the range \mbox{$1.2\le m_c \le 1.8$~GeV}, and
the renormalization scale uncertainty, \mbox{$\mt/2\le\mur\le 2\mt$}.
The renormalization scale has been varied independently for the
direct and resolved contributions. We observe that the central NLO
prediction underestimates the experimental data, which seems to suggest
that the input parameters should be taken such as to yield the largest
possible cross section. This issue will be discussed in more 
detail in the next subsection.

To provide more detailed information on the NLO QCD cross section
prediction and on the theoretical uncertainties, we have collected the
results for the total and the visible cross section in
table~\ref{tab:dir}, including the variation of different input
parameters. We consider the effect of changing the renormalization
scale and the charm quark mass, and of adopting the photon density sets
GRV-HO~\cite{GRV-92} and AFG-HO~\cite{AFG-94}. We also show the
sensitivity to $\Lambda_{\rm QCD}$ by choosing
$\Lambda^{(4)}_{\overline{\rm MS}} = 200$~MeV, which is equal to the
value associated with GRV-HO and is smaller than that associated with
GRS-HO.  The first two columns of the table present total cross
sections in the full phase space (i.e.\ no final-state kinematical cuts are
applied); they differ in the choice of the reference scale $\mu_0$, 
which in the first column
is set to a fixed value (the charm mass), while in the second column
it depends on the final-state kinematics (the scale is set equal to the
transverse mass).  Since the total cross section is dominated by small
transverse momenta, $\pt\simeq m_c$, the difference between
choosing a fixed or a $\pt$-dependent scale is less than about 10\%
for the sum of direct and resolved contributions.  Finally, the third
column displays the predictions for the visible range, defined by the
OPAL cuts. It is obvious from table~\ref{tab:dir} that the visible
cross section is much more stable under scale and mass variations than
the fully extrapolated one and thus provides an important and in
certain aspects better quantity for comparison of data with theory.
The dependence of the theoretical uncertainties upon the kinematical
cuts will also be studied in the following.

\renewcommand{\arraystretch}{1.4}
\begin{table}
\begin{center}
\begin{tabular}{l l c c c } \hline
 & &  &  & $\mu_0=\mt$ \\[-2mm]
$\sigma_{D\bar{D}^*}$ (pb) & & \hspace*{5mm} $\mu_0=m_c$\hspace*{5mm} 
& \hspace*{5mm} $\mu_0=\mt$ 
\hspace*{5mm} & $2<\pt<12$~GeV \\[-2mm]
& & & & $\abs{\eta}<1.5$; $\epsilon=0.035$\\[2mm]
\hline
Total    & Default             & 871.1 & 827.8 & 37.22 \\
         & $\mur=2\mu_0$       & 742.2 & 727.1 & 35.46  \\
         & $\mur=\mu_0/2$      & 1204 & 1035 & 40.45 \\
         & $m_c=1.2$~GeV       & 1712 & 1591 & 41.70 \\
         & $m_c=1.8$~GeV       & 504.8 & 485.6 & 32.7  \\
         & $\Lambda^{(4)}=200$~MeV & 751.2 & 733.0 & 35.35  \\
         & GRV-HO              & 1053 & 984.6 & 42.14 \\
         & AFG-HO              & 906.4 & 856.2 & 39.74 \\ \hline\hline
Direct   & Default             & 590.5 & 577.5 & 26.20 \\
         & $\mur=2\mu_0$       & 528.9 & 522.8 & 26.60  \\
         & $\mur=\mu_0/2$      & 772.7 & 723.0 & 25.57 \\
         & $m_c=1.2$~GeV       & 1098 & 1066 & 27.79 \\
         & $m_c=1.8$~GeV       & 357.3 & 350.9 & 24.10 \\
         & $\Lambda^{(4)}=200$~MeV & 546.0 & 538.4 & 26.46 \\ \hline
Resolved & Default             & 280.6 & 250.3 & 11.02 \\
         & $\mur=2\mu_0$       & 213.3 & 204.3 & 8.860 \\
         & $\mur=\mu_0/2$      & 430.8 & 311.5 & 14.88 \\
         & $m_c=1.2$~GeV       & 613.9 & 524.5 & 13.91 \\
         & $m_c=1.8$~GeV       & 147.5 & 134.7 & 8.659 \\
         & $\Lambda^{(4)}=200$~MeV & 205.2 & 194.6 & 8.894  \\
         & GRV-HO              & 462.6 & 407.1 & 15.94 \\
         & AFG-HO              & 315.9 & 278.7 & 13.54 \\ \hline
\end{tabular} 
\end{center}                                                            
\ccaption{}{\label{tab:dir}
Results for the total cross section broken down to the direct and the
(single- plus double-) resolved photon contributions. The results in
the first two columns are independent of the value of $\epsilon$. The
factorization scale is $\muf=2\mu_0$. The renormalization scale is
$\mur=\mu_0$, unless otherwise indicated. Note that the fragmentation
function is normalized according to $\int_0^1 dz D(z,\epsilon)=1$ as
described in the text.}
\end{table}                                                               

\subsection{DIFFERENTIAL DISTRIBUTIONS}

The study of differential distributions constitutes a more
comprehensive test of the theory than that of total cross sections alone.
In this section we concentrate on the single-inclusive
distributions in transverse momentum $\pt$ and pseudorapidity $\eta$,
for which experimental data are now available.  We emphasize that
with our Monte Carlo programs also more exclusive observables can be
calculated at NLO, so that additional comparisons can be made, should
experimental data become available in the future. In the following, we
have defined the Born cross section for the resolved channels as the
lowest-order partonic cross sections computed with two-loop $\as$ and
convoluted with NLO-evolved photonic parton density sets. This choice 
implies that the difference between the Born and NLO results is essentially
due to the short-distance parton dynamics, as is the case of the direct part.

\begin{figure}
\centerline{
   \epsfig{figure=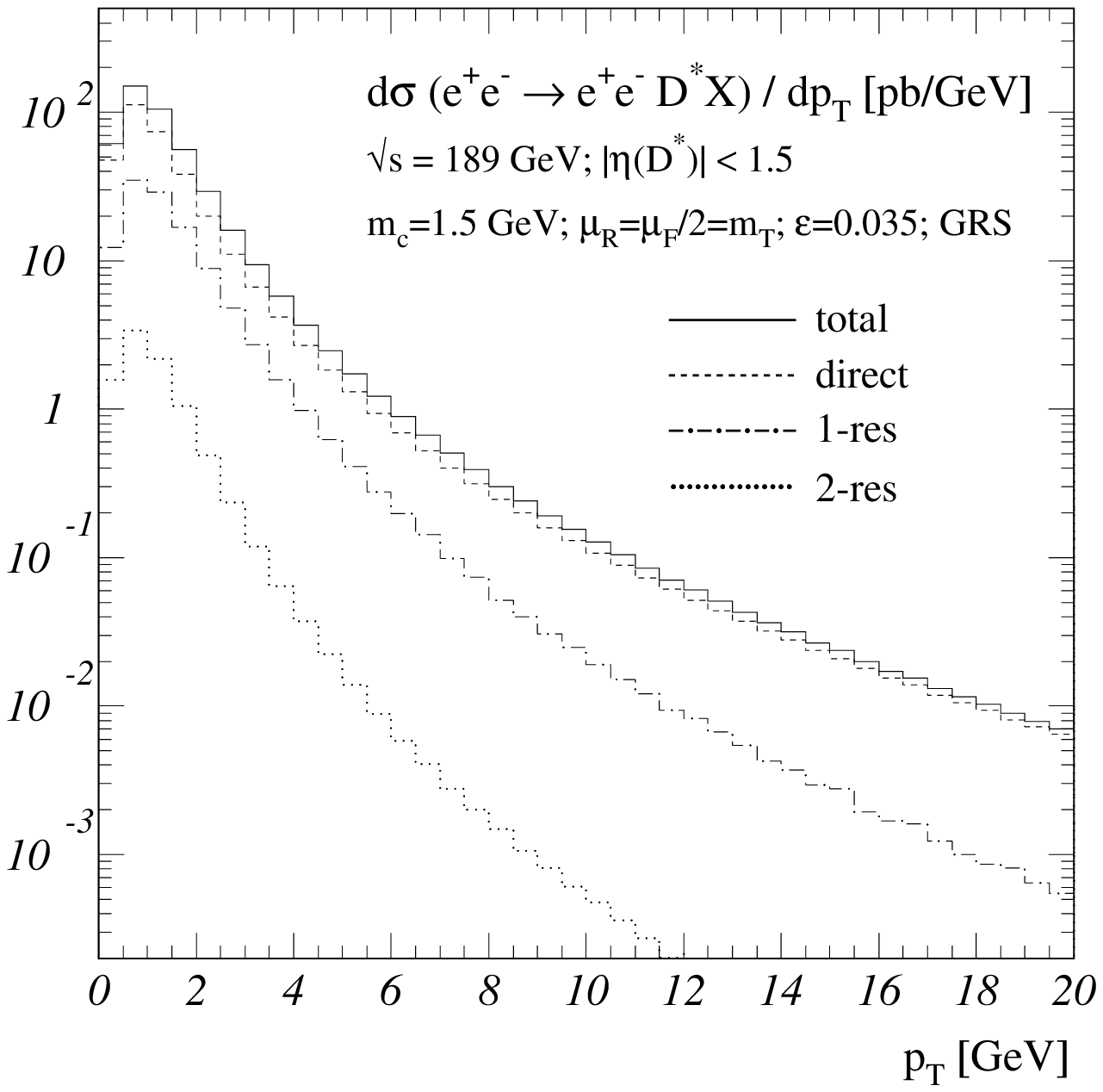,width=0.52\textwidth,clip=}
   \hfill
   \epsfig{figure=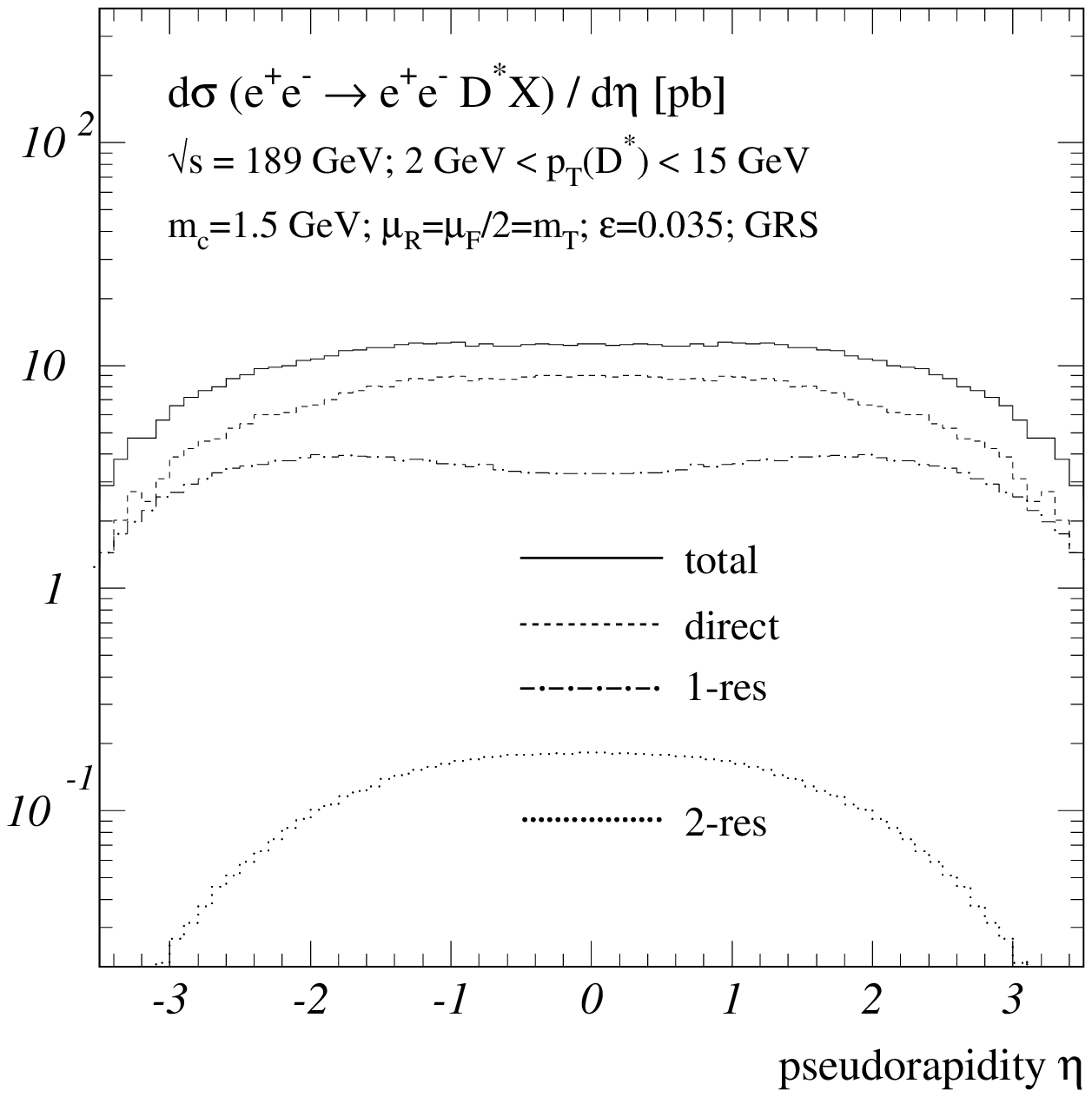,width=0.52\textwidth,clip=} }
\ccaption{}{ \label{fig:spectra}
NLO predictions for the $D^*$ transverse momentum and pseudorapidity
distributions, with default parameters (solid line). The direct,
single-resolved (1-res) and double-resolved (2-res) results are also
shown.  
}
\end{figure}

In fig.~\ref{fig:spectra} we show our predictions for the $\pt$ and
$\eta$ distributions using the default set of parameters. The cuts
$\abs{\eta(D^*)}<1.5$ and $2<\pt(D^*)<15$~GeV have been applied,
respectively. The total cross sections are decomposed into the direct,
single-resolved and double-resolved components. We stress again that
these quantities are not physical in general, although at this order
the direct component is well defined as such. The relative size of the
single- and double-resolved components is factorization-scheme- and
scale-dependent; however, for both the $\overline{\rm MS}$ and
\mbox{DIS$_\gamma$} schemes, and regardless of the specific choice for $\muf$,
the double-resolved component contributes only very marginally to the
charm cross section at LEP2 energies. The direct channel dominates the
cross section in the visible range, even more so for large $\pt$.  As
for the $\eta$ distribution, one observes that the direct channel
dominates at central values, but that the single-resolved contribution
extends further out in pseudorapidity. Therefore, when only large
values of $\eta$ are considered, there is an enhanced sensitivity to
resolved contributions, making it in principle a good observable for
learning about the partonic densities~\footnote{Similar observables
for inclusive jet final states at the Tevatron are indeed used for
such purposes.}.  However, in practice it is very hard for the LEP
experiments to properly identify charmed mesons in such forward
regions. Therefore, we shall not pursue the study of such observables
here.

Let us briefly comment on the effects of resumming large logarithms
$\log (\pt/m_c)$, which arise beyond LO from the collinear emission of
gluons by heavy quarks at large $\pt$, or from almost collinear
branching of gluons into charm-quark pairs. At $\pt\gg m_c$ these
terms might spoil the convergence of the perturbation series and lead
to large scale dependences of the NLO result. The large logarithms can
be resummed by neglecting in the matrix elements all the
power-suppressed mass terms (which behave like $m_c/\pt$ to some
power; this is the reason why this approach is often -- not very
rigorously -- denoted as ``massless''), and absorbing the mass
singularities, occurring as powers of $\log (\pt/m_c)$, into parton
distribution functions and parton-to-heavy-quark fragmentation
functions. The fragmentation functions can be computed in perturbation
theory~\cite{MN-91} at a scale of the order of the heavy quark mass,
and subsequently evolved to the appropriate scale (which is of the
order of $\pt$) by using the Altarelli--Parisi equations; in this
evolution, large logarithms are properly resummed. A study of such a
resummation in photon--photon collisions has been presented in
ref.~\cite{CGKKKS-96}.  We would like to emphasize that the resummed
calculation is not reliable unless $\pt\gg m_c$, since all
non-singular mass terms are neglected.  We have verified numerically
that omitting those mass terms from the NLO cross section
overestimates the full result by no less than $\approx 100\%$ at
$m_c/\pt \approx 1$ and still by $\approx 25\%$ at $m_c/\pt \approx
1/4$. The observation of ref.~\cite{CGKKKS-96}, that resummed and
fixed-order calculation agree down to relatively small values of
$\pt$, must therefore be considered accidental. Similar findings have
been made in an analysis of the $\pt$ spectrum in heavy-flavour
hadroproduction~\cite{CGN-98}. We conclude that a comparison of the
photon--photon $D^*$ data and the resummed approach is not meaningful
unless $\pt(c)\simgt$~6--7~GeV.

\begin{figure}
\centerline{
   \epsfig{figure=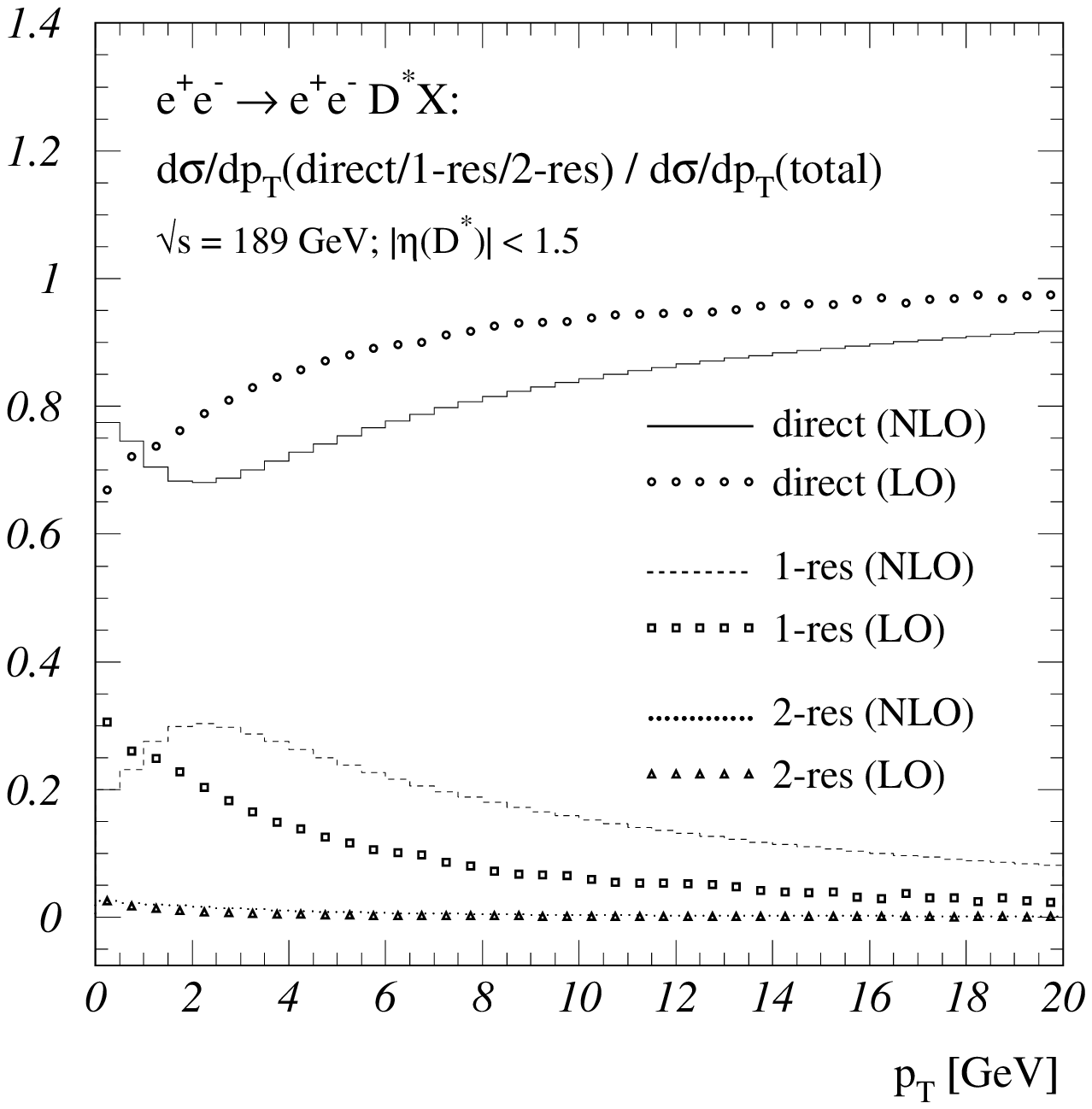,width=0.52\textwidth,clip=}
   \hfill
   \epsfig{figure=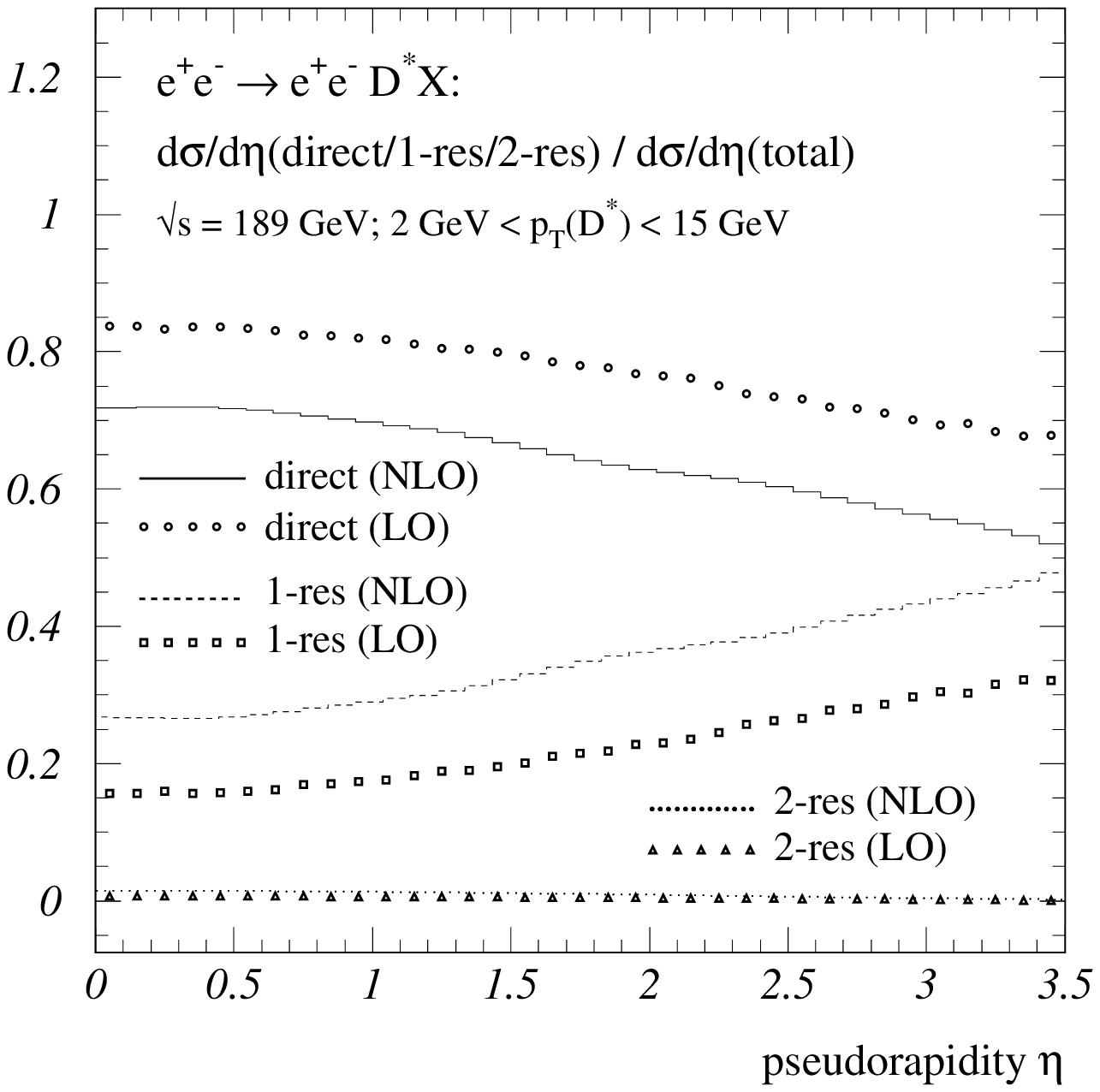,width=0.52\textwidth,clip=} }
\ccaption{}{ \label{fig:relative}
Relative contributions of the direct, single-resolved and
double-resolved processes to the total cross section, as a function of
transverse momentum and pseudorapidity. Both the NLO (histograms) and
Born (symbols) results are presented. Default parameters as specified 
in the text.}
\end{figure}

The relative contribution of the direct, single- and double-resolved
components to the differential cross section is displayed in
fig.~\ref{fig:relative} as a function of $\pt$ and $\eta$.  The curves
are shown for both Born and NLO cross sections. In most of the $\pt$
range the relative fractions at the Born and NLO level are rather
similar, even more so for large $\pt$, since the value of $\as(\mt)$
is decreasing with increasing transverse momentum. On the other hand,
the relative fraction changes sizeably over the whole range of $\eta$
when radiative corrections are included, because the visible cross
section is dominated by the region of low $\pt\;\simgt\; 2$~GeV. In
general, the radiative corrections tend to increase the importance of
the resolved processes. Although the relative contribution of the
resolved cross section never exceeds 30\% (with the default set of
parameters), NNLO corrections may increase the importance of the
resolved channels and thereby enhance the total $D^*$ cross section.
The NLO single-resolved fraction falls below the LO value at low
$\pt\;\simlt\;1$~GeV, causing the NLO direct fraction to exceed the LO
one.  This is due to the $\gamma q$ subprocess in the single-resolved
component, which is only present from NLO onwards and which gives a
negative contribution at low $\pt$. The same behaviour turns out to
occur for the double-resolved component, where the $qg$-initiated
subprocess decreases the NLO cross section at very small $\pt$.

\begin{figure}
\centerline{
   \epsfig{figure=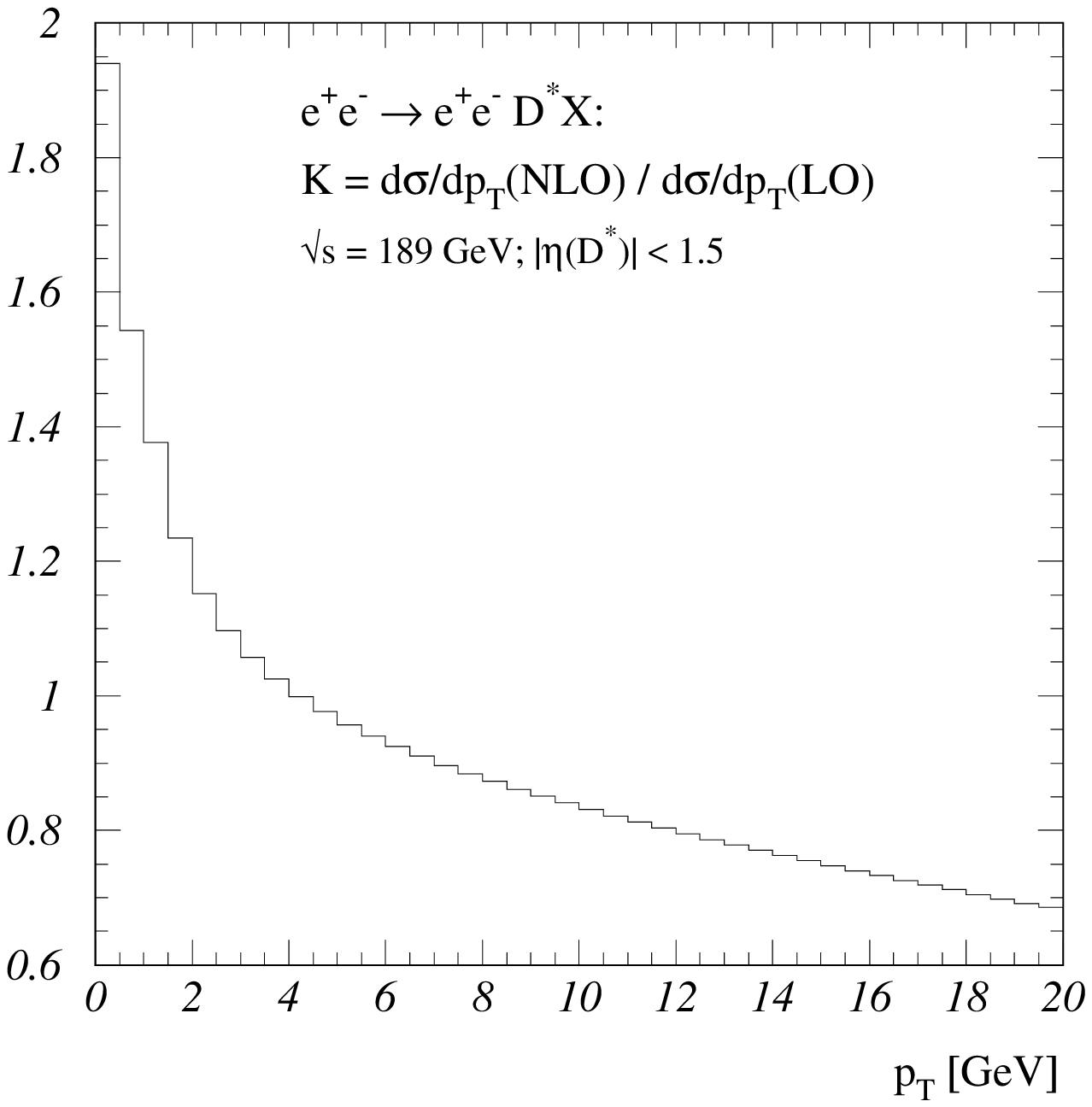,width=0.52\textwidth,clip=} \hfill
   \epsfig{figure=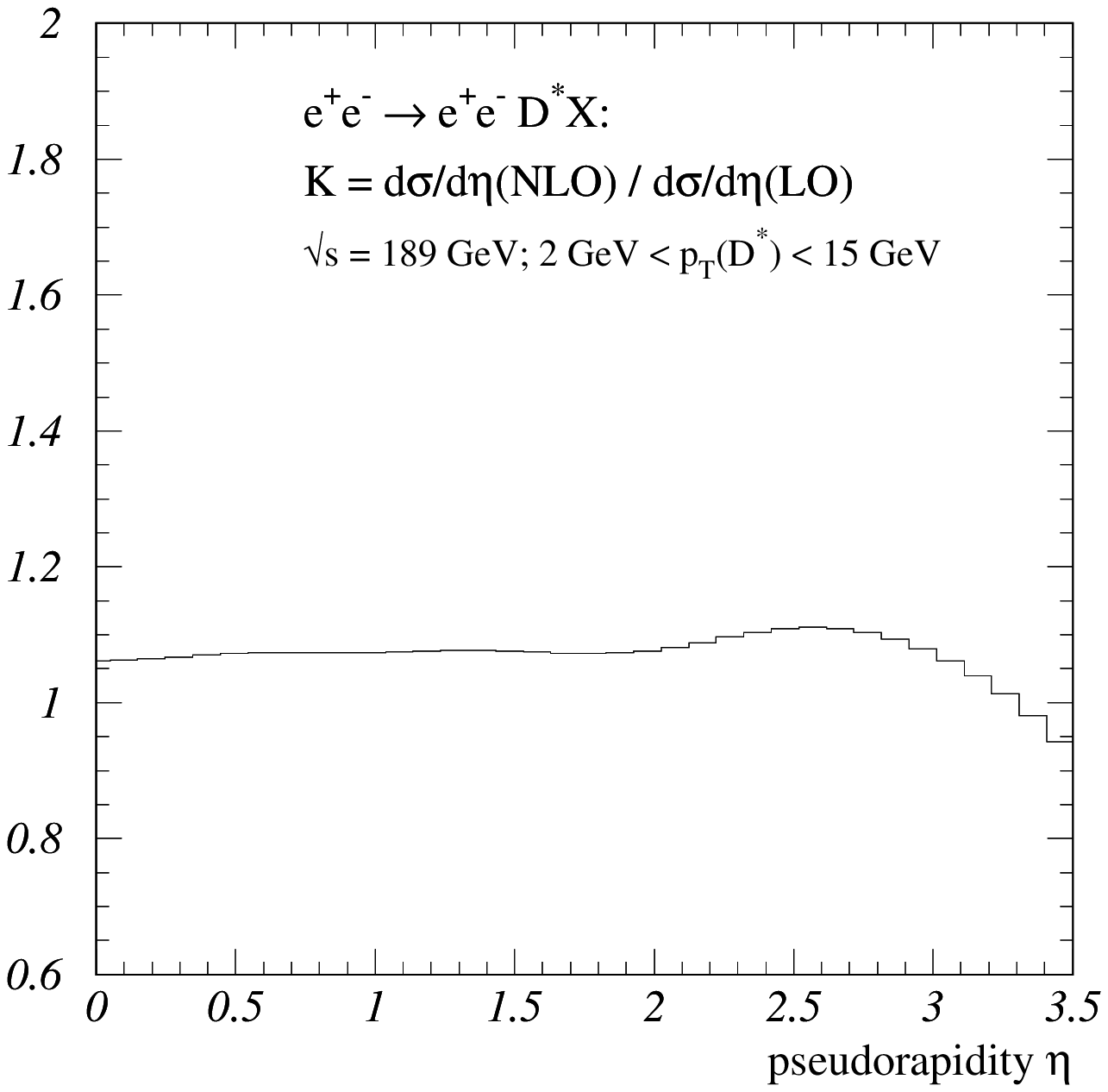,width=0.52\textwidth,clip=} }
\ccaption{}{ \label{fig:kfact}
Ratio of NLO over Born result, as a function of transverse momentum
and pseudorapidity. Default parameters as specified in the text.}
\end{figure}

In fig.~\ref{fig:kfact} we show the ratio of the NLO cross section
over the Born cross section ($K$-factor), as a function of $\pt$ and
$\eta$. The $K$-factor is decreasing with increasing transverse
momentum, a consequence of the decreasing $\as(\mt)$ and the
increasing relative size of the direct contribution. In the visible
range $2<\pt <15$~GeV the $K$-factor is
moderate, ranging approximately from 1.2 to 0.75. This
has to be compared with the case of photon--hadron and hadron--hadron 
collisions, where much larger $K$-factors (up to $K\sim 2$) are observed 
and where NNLO corrections are expected to be sizeable. The dependence 
of the $K$-factor on the pseudorapidity $\eta$, also shown in 
\ref{fig:kfact}, is very mild.

\begin{figure}[H]
\centerline{
   \epsfig{figure=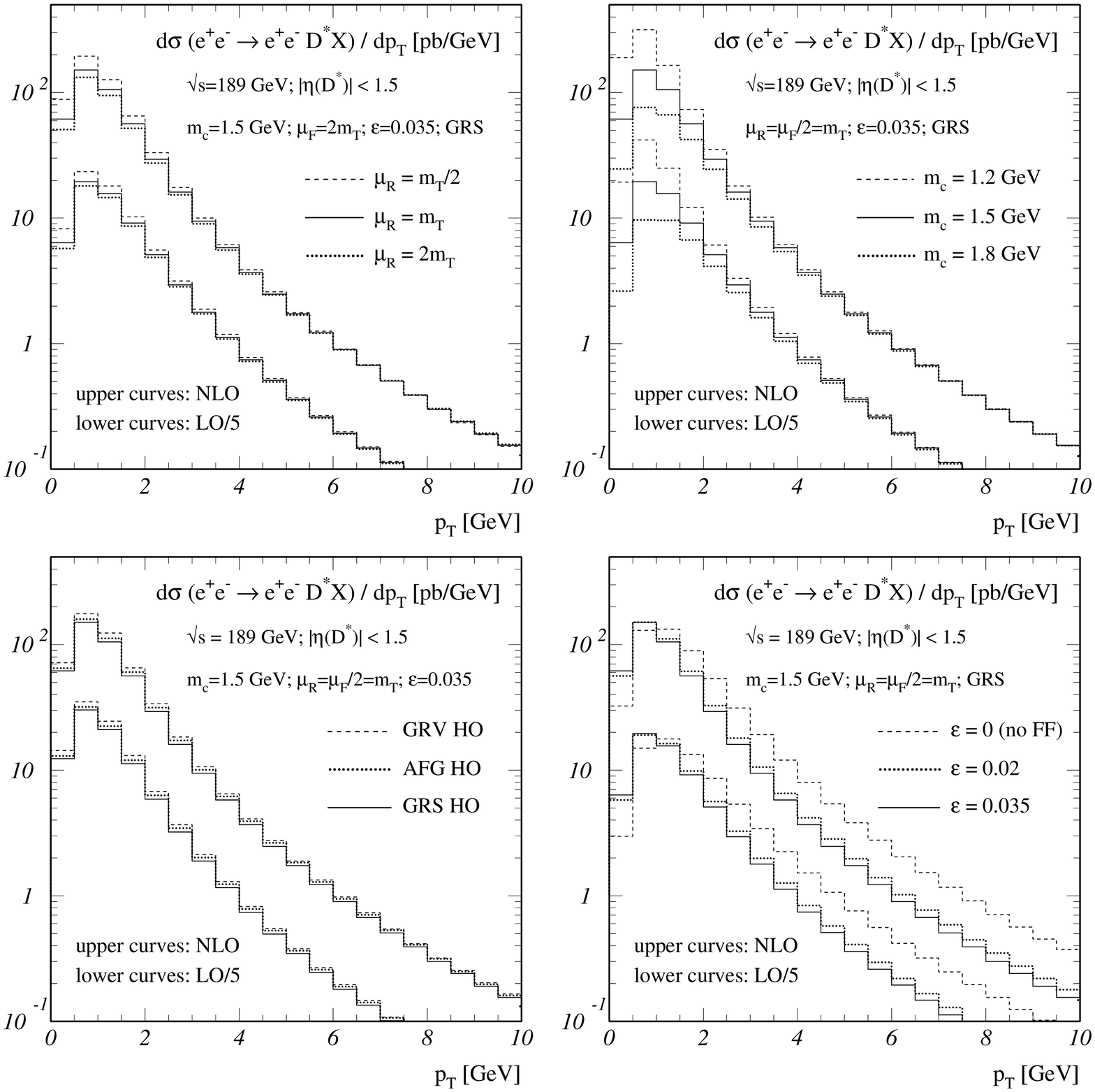,width=1.04\textwidth,clip=}
           }
\ccaption{}{ \label{fig:variation}
Dependence of the $\pt$ spectrum of the $D^*$ upon renormalization scale, 
charm mass, photonic parton densities and the parameter $\epsilon$ of the
Peterson et al.\ fragmentation function. Both the NLO (upper curves)
and the LO (rescaled by 1/5; lower curves) results are presented.}
\end{figure}

Let us now turn to the sensitivity of the transverse momentum
distribution to variations of the input parameters.
Figure.~\ref{fig:variation} shows the $\pt$ spectrum of the $D^*$ for
different choices for the renormalization scale, the charm quark mass,
the NLO photonic parton density set, and the shape of the Peterson et
al.\ fragmentation function as controlled by the parameter
$\epsilon$. The scale, mass and parton density dependence is sizeable
at small $\pt\;\simlt\;2$~GeV, the biggest uncertainty coming from the
variation of the charm quark mass. The strong renormalization scale
and charm mass dependence is mainly induced by the large variation in
the value of $\as(\mt)$. For the lowest charm quark mass value
considered, and at small $\pt$, the hard scale of the process, $\mt
\sim 1.5$~GeV, may indeed be too small to allow for a reliable
perturbative analysis. However, the situation is much improved in the
region probed by experiment, $\pt\;\simgt\;2$~GeV, where the scale and
mass dependence is modest and the theoretical predictions appears to
be well under control.  The reduced theoretical uncertainty is a
consequence both of the increase of the hard scale and of the
increasing importance of the direct contribution, which receives only
small radiative corrections above $\pt\;\simgt\;2$~GeV.
This fact, together with the behaviour of the $K$-factor previously
discussed, leads us to the conclusion that charm production in
photon--photon collisions is under better theoretical
control than in photon--hadron or hadron--hadron collisions.

The dependence upon the parton density sets is mild and visible in the
low-$\pt$ region only, where the resolved contribution is still
sizeable (this is also illustrated in table~\ref{tab:dir}).  The
average Bjorken-$x$ probed is of the order of
$2\mt/\sqrt{s_{\gamma\gamma}}$, so that the photonic parton densities
can in principle be tested in a region where they are poorly
constrained by available data. Unfortunately, the sizeable theoretical
uncertainty due to other sources in the cross section prediction at
small $\pt$ prevents a discrimination of the different PDF sets.

Finally, in the last plot of fig.~\ref{fig:variation} we show the
sensitivity of the $\pt$ spectrum to the shape of the fragmentation
function by varying the $\ep$ parameter that enters the Peterson et
al.\ form.  We have considered, together with our default value
$\ep=0.035$, the more extreme choice $\ep=0.02$, as obtained in recent
fits~\cite{NO-99} where the resummation of large collinear logarithms
is taken into account. We remind the reader that the $\ep$ values
obtained from older fits~\cite{Chrin-87} were significantly higher
($\ep\sim 0.06$), resulting in a softer $\pt$ spectrum. There seems to
be a clear indication now that the degradation of the momentum of the
parent charm quark should be smaller (the very same effect is also
emerging in $b$ physics).  We also display the purely perturbative
result without fragmentation, corresponding to $\ep=0$. \hfill
\noindent
From the figure one can conclude that the $\pt$ spectrum of the $D^*$
in $\gamma\gamma$ collisions is not sensitive enough to the value of
$\ep$ to discriminate between $\ep = 0.035$ and $\ep=0.02$. On the
other hand, the purely perturbative prediction without fragmentation
function results in a much harder spectrum, which, as will be shown
later in this section, is not favoured by the experimental data. Let
us mention that the need to include a fragmentation function is not
always obvious from a comparison with data; measurements of $B$ meson
production in $p\bar{p}$ collisions at the Tevatron are consistent
with a purely perturbative prediction, thus (provocatively) suggesting
$\ep=0$.

\begin{figure}
\centerline{
   \epsfig{figure=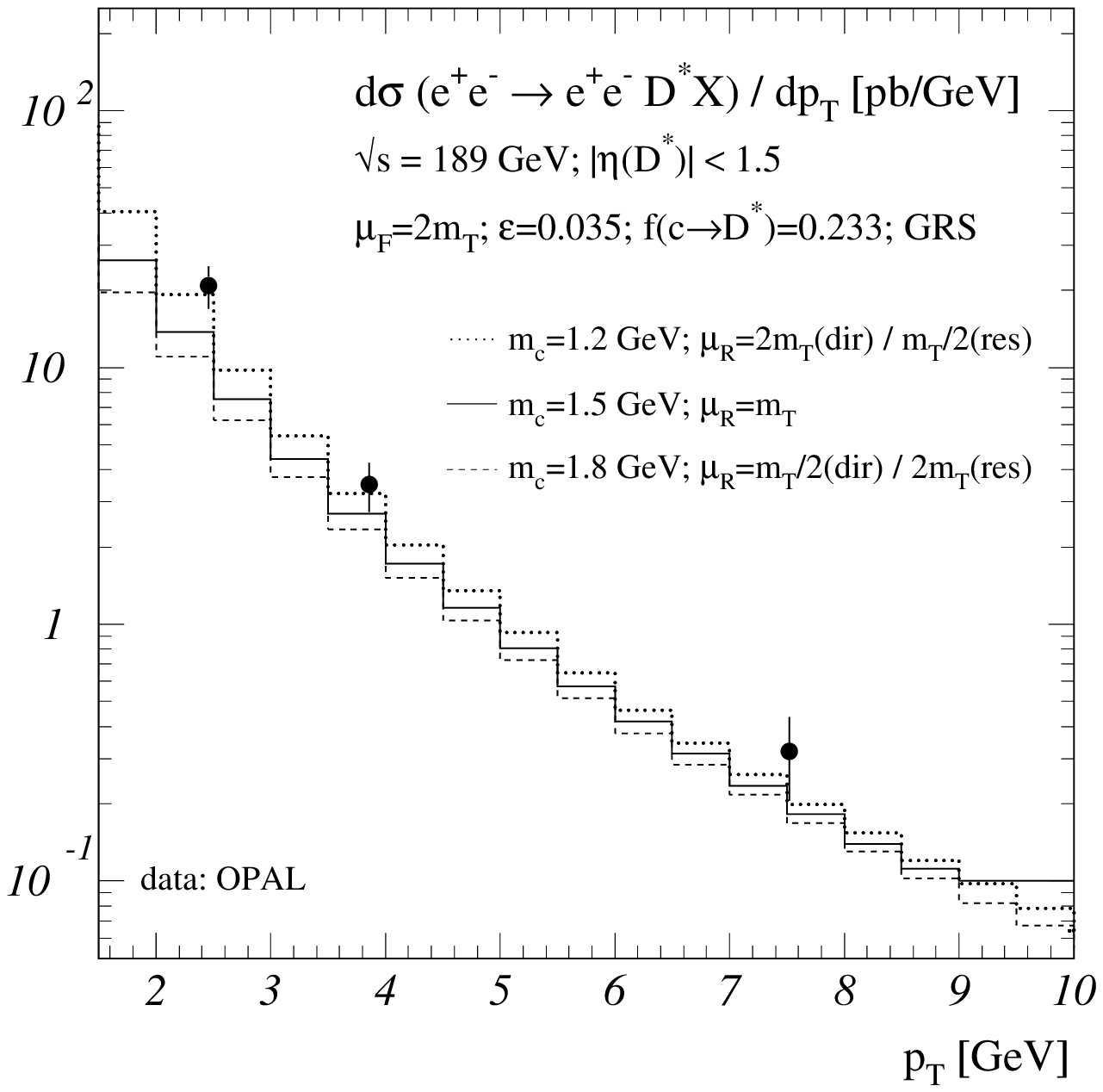,width=0.52\textwidth,clip=}
   \hfill
   \epsfig{figure=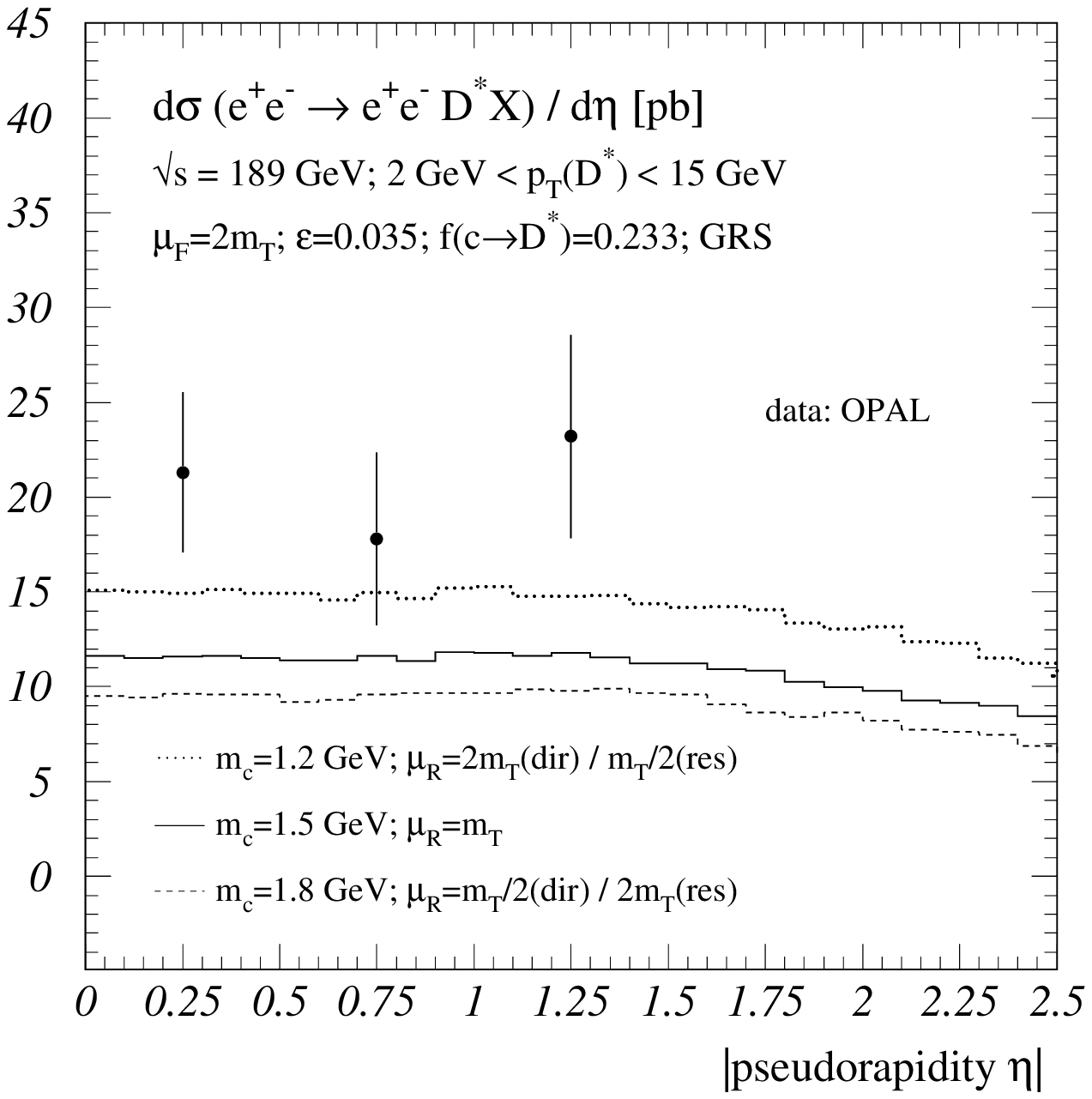,width=0.52\textwidth,clip=} }
\ccaption{}{ \label{fig:comp-data}
Comparison between the NLO theoretical prediction and the OPAL data.
The theoretical band is obtained by varying both the renormalization
scale and the charm mass. The theoretical curves include the
probability for a charm quark to fragment into a $D^*$ meson $f(c\to
D^{*+})=0.233$ and take into account the fact that OPAL counts both
the heavy quark and the heavy antiquark in their cross-section
determination.}
\end{figure}                                                              
The detailed studies of charm production at fixed-target experiments
imply the relevance of non-perturbative effects, such as the intrinsic
transverse momentum of the incoming partons \cite{FMNR-97}.  We
attempted to model such effects by supplying the initial-state partons
with a Gaussian-distributed transverse momentum. Although such a
procedure is not fully consistent when radiative corrections are
included, it can nevertheless provide an estimate of the relevance of
these non-perturbative phenomena.  We have studied this issue also in
the case of photon--photon collisions and find that intrinsic
transverse momentum effects are completely negligible, even if the
average transverse momentum of the partons is chosen as large as 2 to
3 GeV.

In fig.~\ref{fig:comp-data} we finally compare our predictions for the
$D^*$ transverse momentum and pseudorapidity distribution with the
OPAL measurement. Three different theoretical curves are shown, where
the charm quark mass and the renormalization scale are varied as
indicated in the figure, while the photonic parton density and
charm-to-$D^*$ fragmentation function are kept fixed at the default
choice. The theoretical results include the probability for a charm
quark to fragment into a $D^*$ meson $f(c\to D^{*+})=0.233$ and take
into account the fact that OPAL counts both the heavy quark and the
heavy antiquark in their cross-section determination.  We can conclude
that the data agree reasonably well with the NLO predictions.  More
specifically, we can see that the shape of the distributions is
reproduced quite well, while there is a small discrepancy in absolute
normalization, of the same size as that observed when discussing the
total rates. Of course, a definite conclusion will only be possible
after the statistical errors affecting the data will decrease; at
present, the data seem to suggest that the input parameters should be
taken such as to yield the largest possible cross section. However, it
is difficult to disentangle the various effects that could play a
r\^ole here.  We could indeed enlarge the cross section by taking a
small mass value, a small renormalization scale, photonic densities
with much softer gluon, or a combination of these three
effects. Furthermore, the inclusion of NNLO is expected to further
increase our prediction.  In general, we can observe that theory
undershoots the data when central values for the input parameters are
adopted, as is the case for many such comparisons for charm production
at various colliders.

\section{SUMMARY}

We have performed a full NLO QCD study of total and differential $D^*$
production rates in two-photon collisions at LEP2. Compared with
charmed-meson production in hadron--hadron or photon--hadron
collisions, the two-photon cross section appears to be under better
theoretical control, mainly because of the dominance of the direct
channel, where QCD corrections are smaller than in the case of the
resolved channels. This consideration holds in particular for the
moderate- and high-$\pt$ region and for the central $\eta$ region,
which are directly probed by LEP experiments.  We find the largest
theoretical uncertainties, due to charm mass and renormalization scale
dependence, at small $\pt$'s, similarly to what has been observed in
photon--hadron and hadron--hadron collisions.  In this region, the
contribution of the resolved channels is sizeable, and the theoretical
predictions are mildly dependent upon the choice of the photonic
parton densities. However, this dependence is smaller than that due to
mass or renormalization scale choice, and this prevents a
discrimination of the different density sets, even in the ideal case
in which the charm mass would be known exactly. As far as the
comparison with the data available at present is concerned, we find a
good agreement in the shape of the $\pt$ and $\eta$ distributions, and
a reasonable agreement as far as the absolute normalization is
concerned.  In general, the comparison between NLO QCD and the
experimental results reproduces the pattern already known from other
types of collisions; namely, theory somewhat underestimates the data,
and a special tuning of the input parameters is needed to improve the
agreement.  The mild dependence of the theoretical predictions for the
visible cross section upon the input parameters appears to imply that
it will be difficult to constrain any of these parameters by comparing
theory with data. On the other hand, when the statistical significance
of the measurements will be increased, charm production in
photon--photon collisions will be a valuable tool in testing the
underlying production dynamics.

\vspace*{2mm}
\noindent
{\large \bf Acknowledgements}

\noindent
We would like to thank Jochen Patt and Stefan S\"oldner-Rembold for
their enthusiastic encouragement and for providing us with the OPAL
data. We are grateful to Valeri Andreev, Alex Finch, Albert de Roeck,
Gueorgui Soultanov and Andreas Vogt for valuable discussions. The work
of S.F.\ and M.K.\ is supported in part by the EU Fourth Framework
Programme `Training and Mobility of Researchers', Network `Quantum
Chromodynamics and the Deep Structure of Elementary Particles',
contract FMRX-CT98-0194 (DG 12 - MIHT). The work of E.L.\ is part of
the research program of the Foundation for Fundamental Research of
Matter (FOM) and the National Organization for Scientific Research
(NWO).

\section*{APPENDIX: PHOTON--PHOTON CROSS SECTIONS}

In this appendix we describe some technical aspects of the 
computation of charm cross sections in photon--photon collisions.
In particular, we will discuss in some detail the issue of scheme 
dependence of the direct and resolved production mechanisms.

We consider the process
\beq
\begin{array}{r c l}
e^+ + e^- & \longrightarrow & e^+ + e^- + \underbrace{\gamma + \gamma} \\
 &  & \phantom{e^+ + e^- + \gamma\:}\bentarrow Q+\overline{Q}.
\end{array}
\eeq
The final-state positron and electron are scattered almost
collinearly to the beam line; thus, the photons eventually
producing the heavy quark pair are (almost)
on-shell. The cross section can therefore be written as follows
\beq
d\sigma_{ee}(P_{e^+},P_{e^-})=\int dx_{e^+} dx_{e^-}
f^{(e)}_\ga(x_{e^+}) f^{(e)}_\ga(x_{e^-})
d\hat{\sigma}(x_{e^+}P_{e^+},x_{e^-}P_{e^-}),
\label{eexsec}
\eeq
where $d\hat{\sigma}$ is the cross section for the 
process \mbox{$\ga\ga\to Q\bar{Q}$}, and \mbox{$f^{(e)}_\ga$}
is the Weizs\"acker--Williams function \cite{WW-34}. 
We introduce the short-hand notation 
\beq
d\sigma_{ee}=
f^{(e)}_\ga\ast d\hat{\sigma}\ast f^{(e)}_\ga ,
\eeq
which defines the $\ast$ symbol. As mentioned in section 2, 
three mechanisms contribute to the two-photon cross section:
\beq
d\hat{\sigma}=d\hat{\sigma}_{\ga\ga}+d\hat{\sigma}_{1r}
+d\hat{\sigma}_{2r},
\label{gagaxsec}
\eeq
where the three terms in the RHS of this equation denote the direct,
single-resolved, and double-resolved components, respectively.  Using
the factorization theorems in QCD, the single- and double-resolved
cross sections can be written as the convolution of the partonic
hard-scattering cross sections \mbox{$d\hat{\sigma}_{i\ga}$},
\mbox{$d\hat{\sigma}_{\ga i}$} and \mbox{$d\hat{\sigma}_{ij}$}
with the parton densities in the photon \mbox{$f^{(\ga)}_i$}:
\bea
d\hat{\sigma}_{1r}&=&f^{(\ga)}_i\ast d\hat{\sigma}_{i\ga}
+d\hat{\sigma}_{\ga i}\ast f^{(\ga)}_i,
\label{sigores}
\\
d\hat{\sigma}_{2r}&=&f^{(\ga)}_i\ast 
d\hat{\sigma}_{ij}\ast f^{(\ga)}_j,
\label{sigtres}
\eea
Here, a summation over repeated indices is understood, and the indices
$i,j$ run over all the parton flavours $u,\bar{u},\cdots, g$. We
can summarize the content of eqs.~(\ref{gagaxsec}), (\ref{sigores})
and~(\ref{sigtres}) by writing
\beq
d\hat{\sigma}=F^{(\ga)}_a\ast 
d\hat{\sigma}_{ab}\ast F^{(\ga)}_b,
\label{factth}
\eeq
where the indices $a$ and $b$ run over all parton flavours, plus the
photon.  The generalized densities in the photon are given by
\beq
F^{(\ga)}_a(x)=\delta_{a\ga}\delta(1-x)
+f^{(\ga)}_a(x)\left(1-\delta_{a\ga}\right).
\eeq
The three quantities on the RHS of eq.~(\ref{gagaxsec}) are separately
divergent order by order in perturbation theory, and scheme-dependent.
The only quantity that is finite and scheme-independent at all orders
is \mbox{$d\hat{\sigma}$}. However,
as mentioned in the introduction, the reaction we study has the special
feature that the scheme dependence of the direct component and of the 
sum of the resolved components only starts beyond NLO. Although that 
still does not make these cross sections physically observable, it does
give us a fairly clear idea of how often the initial photon fluctuates
to a hadronic state at the energies considered. Let us therefore clarify 
the issue of scheme dependence by working out the specifics in some
detail. We write the perturbative expansions of the partonic cross
sections as:
\bea
d\hat{\sigma}_{\ga\ga}&=&
\aem^2\sum_{n=0}^\infty \as^n d\hat{\sigma}_{\ga\ga}^{(n)},
\label{sigdHO}
\\
d\hat{\sigma}_{i\ga}&=&
\aem\as\sum_{n=0}^\infty \as^n d\hat{\sigma}_{i\ga}^{(n)},
\label{sigorHO}
\\
d\hat{\sigma}_{ij}&=&
\as^2\sum_{n=0}^\infty \as^n d\hat{\sigma}_{ij}^{(n)}.
\label{sigtrHO}
\eea
The expansion of \mbox{$d\hat{\sigma}_{\ga i}$} is identical to that 
of \mbox{$d\hat{\sigma}_{i\ga}$} in eq.~(\ref{sigorHO}). In what
follows, we restrict ourselves to the first non-trivial QCD corrections 
(which amounts to retaining the first two terms in the sums of
eqs.~(\ref{sigdHO})--(\ref{sigtrHO})); this is phenomenologically
relevant, since no complete NNLO result is available for heavy-flavour
production. At this order, only (tree-level and loop) $2\to 2$ and 
(tree-level) $2\to 3$ partonic processes contribute to the result;
they are listed in table~\ref{tab:procs} for the three production
channels.
\begin{table}
\begin{center}
\begin{tabular}{l c c c } \hline
~ & Direct & Single-res. & Double-res. \\
\hline
$2\to 2$ & $\gaga\to\QQB$   & $\gamma g\to\QQB$   & $gg\to\QQB$ \\
         &                  &                     & $q\bar{q}\to\QQB$ \\
\hline
$2\to 3$ & $\gaga\to\QQB g$ & $\gamma g\to\QQB g$ & $gg\to\QQB g$ \\
         &                  & $\gamma q\to\QQB q$ & $q\bar{q}\to\QQB g$ \\
         &                  &                     & $qg\to\QQB q$ \\
\hline
\end{tabular} 
\end{center}                                                            
\ccaption{}{\label{tab:procs}
Partonic subprocesses contributing to heavy-quark production
in $\gaga$ collisions at next-to-leading order. 
}
\end{table}                                                               

A general two-photon cross section such as $d\hat{\sigma}$, being a 
physical observable, must be scheme-independent.  Therefore, an 
expression equivalent to eq.~(\ref{factth}) is
\beq
d\hat{\sigma}=F^{(\ga)^\prime}_a\ast 
d\hat{\sigma}_{ab}^\prime\ast F^{(\ga)^\prime}_b,
\label{xsecscheme}
\eeq
where the prime denotes that the generalized densities and the 
partonic cross sections, which are non-physical, are calculated
in a scheme different from that used in eq.~(\ref{factth}).
In QCD, the parton densities expressed in two different schemes
can always be related as follows:
\beq
F^{(\ga)^\prime}_a=\delta_{a\ga}\delta
+f^{(\ga)^\prime}_a\left(1-\delta_{a\ga}\right),\;\;\;\;
f_i^{(\ga)^\prime}=f_i^{(\ga)}+\aem H_i
+\as K_{ij}\otimes f_j^{(\ga)},
\label{Fscheme}
\eeq
where $H_i$ and $K_{ij}$ are functions and distributions,
respectively, with an expansion in $\as$ starting from $\as^0$. The
symbol $\otimes$ denotes the convolution integral
\beq
A\otimes B(z)=B\otimes A(z)=\int dx dy \,\delta(z-xy)\, A(x) B(z).
\eeq
Substituting eq.~(\ref{Fscheme}) into eq.~(\ref{xsecscheme}),
and using eqs.~(\ref{sigdHO})--(\ref{sigtrHO}) we find:
\bea
d\hat{\sigma}&\!\!=\!\!&
\aem^2 d\hat{\sigma}_{\ga\ga}^{(0)^\prime}
+\aem^2\as\left[d\hat{\sigma}_{\ga\ga}^{(1)^\prime}
+H_i\ast d\hat{\sigma}_{i\ga}^{(0)^\prime}
+d\hat{\sigma}_{\ga i}^{(0)^\prime}\ast H_i\right]
\nonumber \\*
&\!\!+\!\!&\aem\as\left[f^{(\ga)}_i\ast d\hat{\sigma}_{i\ga}^{(0)^\prime}
+d\hat{\sigma}_{\ga i}^{(0)^\prime}\ast f^{(\ga)}_i\right]
+\aem\as^2\left[f^{(\ga)}_i\ast d\hat{\sigma}_{i\ga}^{(1)^\prime}
+d\hat{\sigma}_{\ga i}^{(1)^\prime}\ast f^{(\ga)}_i\right.
\nonumber \\*
&\!\!+\!\!&\left. 
\!\!\left(K_{ik}\otimes f_k^{(\ga)}\right)
\ast d\hat{\sigma}_{i\ga}^{(0)^\prime}
+d\hat{\sigma}_{\ga i}^{(0)^\prime}\ast 
\left(K_{ik}\otimes f_k^{(\ga)}\right)
+H_i\ast d\hat{\sigma}_{ij}^{(0)^\prime}\ast f_j^{(\ga)}
+f_i^{(\ga)}\ast d\hat{\sigma}_{ij}^{(0)^\prime}\ast H_j
\right]
\nonumber \\*
&\!\!+\!\!&
\as^2 f^{(\ga)}_i\ast d\hat{\sigma}_{ij}^{(0)^\prime}\ast f^{(\ga)}_j 
+\as^3\left[f^{(\ga)}_i\ast d\hat{\sigma}_{ij}^{(1)^\prime}\ast f^{(\ga)}_j
+f^{(\ga)}_i\ast d\hat{\sigma}_{ij}^{(0)^\prime}\ast 
\left(K_{jk}\otimes f_k^{(\ga)}\right)\right.
\nonumber \\*
&\!\!+\!\!&\left. 
\!\!\left(K_{ik}\otimes f_k^{(\ga)}\right)\ast
d\hat{\sigma}_{ij}^{(0)^\prime}\ast f^{(\ga)}_j\right]
+{\cal O}\left(\aem^2\as^2,\aem\as^3,\as^4\right).
\label{xsecfinal}
\eea
Using only the first term in the expansion in $\as$ of $H_i$ and
$K_{ij}$ (which we denote as $H_i^{(0)}$ and $K_{ij}^{(0)}$,
respectively), and comparing eq.~(\ref{xsecfinal}) with
eq.~(\ref{gagaxsec}), we get
\bea
&&d\hat{\sigma}_{\ga\ga}^{(0)^\prime}=
d\hat{\sigma}_{\ga\ga}^{(0)},\;\;\;\;
d\hat{\sigma}_{i\ga}^{(0)^\prime}=
d\hat{\sigma}_{i\ga}^{(0)},\;\;\;\;
d\hat{\sigma}_{\ga i}^{(0)^\prime}=
d\hat{\sigma}_{\ga i}^{(0)},\;\;\;\;
d\hat{\sigma}_{ij}^{(0)^\prime}=
d\hat{\sigma}_{ij}^{(0)},
\label{pvsnpLO}
\\&&
d\hat{\sigma}_{\ga\ga}^{(1)^\prime}=
d\hat{\sigma}_{\ga\ga}^{(1)}
-H_i^{(0)}\ast d\hat{\sigma}_{i\ga}^{(0)}
-d\hat{\sigma}_{\ga i}^{(0)}\ast H_i^{(0)},
\label{pvsnpd}
\\&&
d\hat{\sigma}_{i\ga}^{(1)^\prime}=
d\hat{\sigma}_{i\ga}^{(1)}
-K_{ki}^{(0)}\ast d\hat{\sigma}_{k\ga}^{(0)}
-d\hat{\sigma}_{ik}^{(0)}\ast H_k^{(0)},
\label{pvsnporo}
\\&&
d\hat{\sigma}_{\ga i}^{(1)^\prime}=
d\hat{\sigma}_{\ga i}^{(1)}
-d\hat{\sigma}_{\ga k}^{(0)}\ast K_{ki}^{(0)}
-H_k^{(0)}\ast d\hat{\sigma}_{ki}^{(0)},
\label{pvsnport}
\\&&
d\hat{\sigma}_{ij}^{(1)^\prime}=
d\hat{\sigma}_{ij}^{(1)}
-d\hat{\sigma}_{ik}^{(0)}\ast K_{kj}^{(0)}
-K_{ki}^{(0)}\ast d\hat{\sigma}_{kj}^{(0)},
\label{pvsnptr}
\eea
where use has been made of the fact that 
\beq \left(A\otimes
B\right)\ast C = A\ast\left(B\ast C\right).  
\eeq
Equation.~(\ref{pvsnpLO}) is nothing but the formal statement that, 
at leading order, the cross sections are independent of the scheme
adopted. This equation also implies that, {\it at this order}, the
direct, single-resolved, and double-resolved terms are separately
physically meaningful. The inclusion of radiative corrections,
however, changes the situation, as can be clearly seen from
eqs.~(\ref{pvsnpd})--(\ref{pvsnptr}).  For instance, when changing 
the scheme, the single-resolved cross section (eqs.~(\ref{pvsnporo})
and~(\ref{pvsnport})) receives a contribution from the double-resolved
part. In principle, this is also true for the direct cross section
(eq.~(\ref{pvsnpd})). Note, however, that for heavy flavour
production, if $i$ is a gluon, then $H_i^{(0)}$ is zero, while if $i$
is a quark, then \mbox{$d\hat{\sigma}_{i\ga}^{(0)}$} is zero
(this is not true for -- say -- jet production,
where \mbox{$d\hat{\sigma}_{q\ga}^{(0)}$} is non-vanishing).
Therefore, at next-to-leading order the direct term is
scheme-independent, while the single- and double-resolved terms are
closely related, since one is entitled to add to them a finite piece
without changing their sum. Clearly, when going to
next-to-next-to-leading order, the direct cross section will also
become scheme-dependent.

\newpage



\end{document}